\documentclass[reprint,superscriptaddress, amsmath,amssymb, aps, pra, longbibliography]{revtex4-1}

\usepackage{soul}
\usepackage{amsmath}   
\usepackage{amssymb}
\usepackage{mathtools}
\usepackage{graphicx}
\usepackage{dcolumn}
\usepackage{bm}
\usepackage{amsmath}
\usepackage{graphicx}
\usepackage{amsfonts}
\usepackage{subfigure}
\usepackage{graphicx}
\usepackage{array}
\usepackage{float}
\usepackage{color}
\usepackage{multirow}
\usepackage{amssymb}%
\usepackage[colorlinks=true,linkcolor=blue]{hyperref}%
\hypersetup{allcolors=blue}
\usepackage[normalem]{ulem}
\usepackage{xcolor}

\newcommand{\ket}[1]{\ensuremath{\left\vert #1 \right\rangle}}
\usepackage{mathtools}
\DeclarePairedDelimiterX\braket[2]{\langle}{\rangle}{#1 \delimsize\vert #2}

\begin{document}
\title{Hyperfine structure and collisions in three-photon Rydberg electromagnetically induced transparency}
\date{\today }

\author{Alisher Duspayev}
    \email{alisherd@umich.edu}
    \thanks{Present address: Department of Physics and Joint Quantum Institute, University of Maryland, College Park, MD 20742, USA}
    \affiliation{Department of Physics, University of Michigan, Ann Arbor, MI 48109, USA}   
\author{Georg Raithel}
    \affiliation{Department of Physics, University of Michigan, Ann Arbor, MI 48109, USA}  

\begin{abstract}
Multi-photon electromagnetically-induced transparency (EIT) of atomic vapors involves several intermediate atomic levels. The sub-structure of these levels and their collisional interactions can drastically alter experimental EIT signals. Here, we report on hyperfine structure and collision effects in three-photon Rydberg EIT on the cascade $5S_{1/2} \rightarrow$ $5P_{1/2} \rightarrow 5D_{3/2}$ $\rightarrow 25F_{5/2}$ in a room temperature $^{85}$Rb vapor cell. In our measurements of EIT spectra, we identify two types of EIT signatures that correspond with distinct excitation pathways and atomic velocity classes in the atomic vapor. The $5D_{3/2}$ hyperfine structure and Autler-Townes splittings lead to complex patterns in the EIT spectra, which we analyze with the aid of 10-level EIT simulations. Adding 50~mTorr of Ar gas alters the EIT spectra and induces an additional, third EIT mode. Based on our simulation results, we attribute these changes to hyperfine collisions in the Rb $5D_{3/2}$ level. Our study may become useful in quantum technologies involving Rydberg EIT and hyperfine collisions in vapor cells, including non-invasive spatio-temporally resolved electric-field sensing of electric fields in low-pressure plasmas.
\end{abstract}

\maketitle

% ----------------------------------------------------------------
\section{Introduction}
\label{sec:intro}

Electromagnetically-induced transparency (EIT)~\cite{boller1991} has become an important technique for modern research in quantum optics~\cite{liang}, quantum information processing~\cite{phillipsprl2001} and Rydberg-atom-based field sensing~\cite{mohapatra2007, Sedlacek2012, Holloway2014}. Focusing on the latter application, EIT offers simple, yet efficient and nondestructive detection of Rydberg states in room-temperature alkali-atom vapors. Although most of the ground-breaking work in these systems has been performed with two-photon schemes, there is an interest to go beyond that~\cite{carr2012, Thaicharoen2019}. Three-photon Rydberg-EIT schemes can be realized with all-infrared lasers~\cite{thoumany2009, Fahey11, johnson2012}, which allows one to cancel Doppler shifts~\cite{bohaichuk2023} and to excite Rydberg states with angular momentum $\ell \ge 3$ in warm vapors~\cite{highellpaper, prajapati2023highell, allinson2024}. All-infrared EIT laser schemes also reduce photo-electric effects and resultant light-induced electric fields in cells~\cite{ma20}. These features could be beneficial for electric field measurements in the VHF and UHF radio bands~\cite{brown2023, elgee2023, prajapati2023highell}, DC electric field sensing~\cite{anderson2017, highellpaper} for applications in plasma physics and technology~\cite{alexiou1995,park2010, Goldberg_2022} as well as microfield diagnostics~\cite{ionsourcepaper} that occur within in-vacuum focused-ion beam sources~\cite{bassim2014recent, gierak, knuffman2013, McCulloch_2016}.  

A typical three-photon Rydberg-EIT cascade scheme involves a stable ground state, two rapidly decaying intermediate states, and a high-lying metastable Rydberg state. These states are coupled by three laser fields. A suitable example in rubidium atom is the cascade system $5S_{1/2}$ - $5P_{J'}$ - $5D_{J''}$ - $nF_{J'''}$ with an $\ell=3$ Rydberg state, with principal quantum number $n$. The hyperfine structures of the intermediate states range from a few tens of MHz to hundreds of MHz, leading to line splittings in the EIT spectra~\cite{Yang2011} and interference of transition amplitudes in excitation channels~\cite{Duspayev2023}. Hyperfine effects lead to increased complexity in EIT spectra~\cite{Badger_2001}. In some cases, it is possible to eliminate the hyperfine effects by using intermediate-state detunings in the EIT cascade system that are much larger than the hyperfine splittings~\cite{Thaicharoen2019, highellpaper}, at the expense of reduced EIT signal strength. Conversely, hyperfine-changing collisions at small intermediate-state detunings introduce additional quantum degrees of freedom within the driven atomic vapor that may become useful~\cite{katz2022}.

Here, we investigate hyperfine-induced structures in the exemplary three-photon-cascade EIT system $5S_{1/2} \rightarrow$ $5P_{1/2} \rightarrow$ $ 5D_{3/2} \rightarrow$ $25F_{5/2}$ in $^{85}$Rb. We obtain EIT spectra on a two-dimensional map spanned by the detuning $\Delta_D$ of the laser driving the $5P_{1/2} \rightarrow 5D_{3/2}$ transition (dressing laser) and the detuning $\Delta_R$ of the laser driving the $5D_{3/2} \rightarrow nF_{5/2}$ Rydberg transition (coupler laser). The collected experimental Rydberg-EIT spectra exhibit a rich structure of hyperfine-induced features that we analyze with the aid of numerical models. We observe and explain several three-photon EIT modes. In the near-resonant region, $\Delta_D \approx 0$ and $|\Delta_R| \lesssim 400~$MHz, we find a set of EIT features that are generated by the hyperfine structure of the intermediate $5D_{3/2}$ level. Data taken in the presence of an additional low-pressure inert gas reveal an extra EIT branch that is attributed to hyperfine collisions of the intermediate levels of the three-photon EIT cascade. The collisional EIT branch provide insight into which types of inert-gas-induced interactions are the most relevant.      

\section{Experimental setup}
\label{sec:setup}

\begin{figure}[t!]
 \centering
  \includegraphics[width=0.45\textwidth]{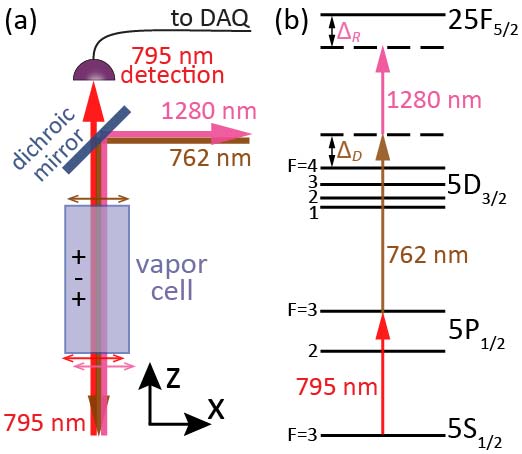}
  \caption{Experimental setup (a) and relevant $^{85}$Rb energy-level diagram (b). All lasers are linearly polarized parallel to the $x$-axis.
  The 795-nm and 1280-nm lasers propagate in $+z$, and the 762-nm laser propagates in $-z$-direction  (``+ - +'' configuration). The diagram in (b) includes the hyperfine structures of the $5P_{1/2}$ and the $5D_{3/2}$ states (not to scale).} 
  \label{fig:setup}
\end{figure}

Our experimental setup is sketched in Fig.~\ref{fig:setup}~(a) with the diagram of the relevant $^{85}$Rb energy levels in Fig.~\ref{fig:setup}~(b). Rydberg atoms are excited in a room-temperature vapor cell using three infrared lasers. The lowest stage, which constitutes the EIT probe, has a wavelength $\lambda_P \approx$~795~nm and is frequency-locked to the $\ket{5S_{1/2}, F = 3} \rightarrow \ket{5P_{1/2}, F' = 3}$ transition using saturated-absorption spectroscopy in a first reference vapor cell. The probe laser has a detuning $\Delta_P \approx 0$ from said transition and is sent to the measurement cell via a polarization-maintaining (PM) fiber. The probe light transmitted through the measurement cell is detected as shown in Fig.~\ref{fig:setup}~(a).

The intermediate excitation laser, which provides the EIT dressing beam, has a wavelength $\lambda_D \approx$~762~nm. The 762-nm laser is split into two beams. One passes through a fiber-coupled electro-optic modulator (EOM) and then through a second reference vapor cell. There it is overlapped with and counter-aligned to a 795-nm beam sample from the frequency-locked probe laser. The EOM has a near-linear response from DC to about 10~GHz and is driven by a tunable radio-frequency (RF) source. The 762-nm beam passing through the second reference cell has three significant components, namely the carrier and the first-order frequency modulation (FM) sidebands, each of which generates a two-photon EIT signal in the second reference cell on the $5S_{1/2} \rightarrow5P_{1/2} \rightarrow 5D_{3/2}$ cascade. The 762-nm laser is frequency-locked to the reference EIT signal on one of the FM sidebands. The carrier is scanned by changing the frequency of the EOM drive. The second, unmodulated 762-nm beam, which has a frequency equal to that of the carrier passing through the second reference vapor cell, is sent via another PM fiber to the measurement cell. The dressing-beam detuning with respect to the $\ket{5P_{1/2}, F' = 3} \rightarrow \ket{5D_{3/2}, F'' = 4}$ hyperfine transition, denoted $\Delta_D$ [see Fig.~\ref{fig:setup}~(b)], is varied via the RF frequency of the EOM driver. The $\Delta_D$ scan range is about 100~MHz wide and is approximately centered to the 30-MHz wide manifold of the optically coupled $^{85}$Rb $5D_{3/2}$ hyperfine states $F''=2$, 3 and 4. 

The upper-stage laser provides the EIT coupler beam, which in the present work drives the $\ket{5D_{3/2}, F''=*} \rightarrow \ket{25F_{5/2}, F'''=*}$ Rydberg transitions. The coupler has a wavelength $\lambda_R \approx$~1280~nm, and its detuning relative to the $\ket{5D_{3/2}, F''=4} \rightarrow \ket{25F_{5/2}, F'''=*}$ hyperfine component is denoted $\Delta_R$. The allowed Rydberg hyperfine states, $F'''=1$ to 5, are not resolved due to their minute, albeit unknown, hyperfine splittings. The detuning $\Delta_R$ is scanned by applying a linear voltage ramp to a piezoelectric tuning element in the 1280-nm laser.  
Most of the 1280-nm power is used as a seed for a fiber-coupled optical amplifier. The amplifier's output beam is sent through an acousto-optic modulator (AOM). A first-order AOM diffraction mode is aligned through the measurement cell, as shown in Fig.~\ref{fig:setup}~(a). The AOM is used to pulse-modulate the coupler beam for lock-in detection of the EIT probe signal. Here, we modulate the coupler with a 1.5~kHz square signal with a 50\% duty cycle. To linearize and calibrate the $\Delta_R$-scans, a coupler-laser beam sample is passed through another EOM that FM-modulates the sample at a frequency of 125~MHz. The FM-modulated beam sample is passed through a Fabry-P\'erot (FP) etalon with a free spectral range (FSR) of $\approx$~375~MHz, and the transmitted sample power is measured with a photodiode. The FP is air-spaced, has a fixed ultra-low-expansion glass spacer, and is housed in a temperature-stabilized fixed-pressure container. As the 1280-nm laser is scanned, the transmission of the 1280-nm beam sample through the FP etalon exhibits peaks at a fixed frequency spacing of 125~MHz. These are recorded simultaneously with the 795-nm EIT-probe transmission through the EIT measurement cell. There are about 6 FP transmission peaks in each $\Delta_R$ scan, which is sufficient to map the EIT coupler scans onto a linear, calibrated $\Delta_R$ frequency axis. The FP transmission peaks also serve to correct for long-term coupler-laser frequency drifts. 

All three lasers are commercial external-cavity diode lasers. As depicted in Fig.~\ref{fig:setup}~(a), the 795-nm and the 1280-nm beams co-propagate along the $+z$-direction in the measurement cell, whereas the 762-nm beam is counter-propagating to them. This configuration is referred to as $(+,-,+)$~\cite{Thaicharoen2019}. All lasers are polarized along the $x$-direction. The optical powers entering the measurement cell are $\approx$~10~$\mu$W, $\approx$~4~mW and $\approx$~10~mW, and the beams are focused down to a full width at half maximum (FWHM) of $\approx$~80~$\mu$m, $\approx$~250~$\mu$m and $> 250~\mu$m for the 795-nm, 762-nm and 1280-nm EIT beams, respectively. Using these beam parameters, we estimate the respective radial Rabi frequencies at the beam centers to be $\Omega_{PS} \approx$~50~MHz for $\ket{5S_{1/2}, F = 3} \rightarrow \ket{5P_{1/2}, F' = 3}$, $\Omega_{DP} \approx$~100~MHz for $\ket{5P_{1/2}, F' = 3} \rightarrow \ket{5D_{3/2}, F''=*}$, and $\Omega_{RD} \lesssim $~25~MHz for the Rydberg transitions. The $5D_{3/2}$ hyperfine splittings are several tens of MHz and are spectroscopically resolved. The unknown Rydberg $25F_{5/2}$ hyperfine structure is very small and cannot be resolved. The radial Rabi frequencies are multiplied with angular matrix elements, as described in Sec.~\ref{sec:theory}. All Rabi frequencies drop off from their center values according to the respective FWHM beam diameters. A dichroic mirror [see Fig.~\ref{fig:setup}~(a)] is used to overlap all beams and to deflect the 1280-nm light from the photodiode that detects the 795-nm EIT probe signal. 

\section{Data acquisition}
\label{sec:acquisition}

The data are acquired in the following order. For a selected value of $\Delta_D$, the detuning of the EIT coupler beam, $\Delta_R$, is scanned over a range of $\approx$~700~MHz with a scan duration of 100~ms. To improve the signal-to-noise ratio of the measured EIT probe signal, the coupler beam is pulse-modulated at 1.5~kHz while $\Delta_R$ is being scanned, and the transmission of the EIT probe beam, $T$, is demodulated using a lock-in amplifier. The lock-in amplifier is referenced to the waveform that square-modulates the coupler beam. The lock-in amplifier's time constant is set at 3~ms. The lock-in amplifier's phase is set once so that the magnitude of the ``X'' output is largest, while the ``Y'' quadrature output is near zero. The ``X'' output constitutes the recorded signal, denoted $S_0(\Delta_R)$. It is proportional to the difference in probe transmission for the coupler laser on ($T$) and off ($T_0$), 
\[
S_0(\Delta_R) \approx const. \times [T(\Delta_R) - T_0(\Delta_R)]  \quad .
\]
Positive $S_0$ corresponds to EIT and negative  $S_0$ to electromagnetically induced absorption (EIA).  As the coupler laser is scanned, $S_0(\Delta_R(t))$ is acquired on an oscilloscope in AC mode (time constant 50~ms). The recorded signal, $S(\Delta_R(t))$, is relative to the time average $\langle S_0(\Delta_R(t)) \rangle_t$,
\[ S(\Delta_R(t)) = S_0(\Delta_R(t)) - \langle S_0(\Delta_R(t)) \rangle_t \quad .\] 
For each value of $\Delta_D$, we acquire 100 to 150 scans $S(\Delta_R(t))$, which are averaged. Using the simultaneously recorded FP transmission peaks as frequency markers, the averaged time-dependent scan is converted into a function of coupler-laser detuning, $\Delta_R$. In this way, the mild non-linearities of the coupler-laser scan are eliminated, and the data $\bar{S}(\Delta_R)$ are rendered on a calibrated, linear $\Delta_R$-scale. The data $\bar{S}(\Delta_R)$ are then shifted by small amounts so that the $\Delta_R$-averages are zero, {\sl{i.e.}} $\langle \bar{S}(\Delta_R) \rangle_{\Delta_R}=0$.
The procedure is repeated for selected values of $\Delta_D$ on an equidistant grid with a step size of 2~MHz. All scans are then assembled in two-parameter maps, $\bar{S}(\Delta_R, \Delta_D)$.

\section{Results without added inert gas}
\label{sec:without}

\begin{figure}[t!]
 \centering
  \includegraphics[width=0.45\textwidth]{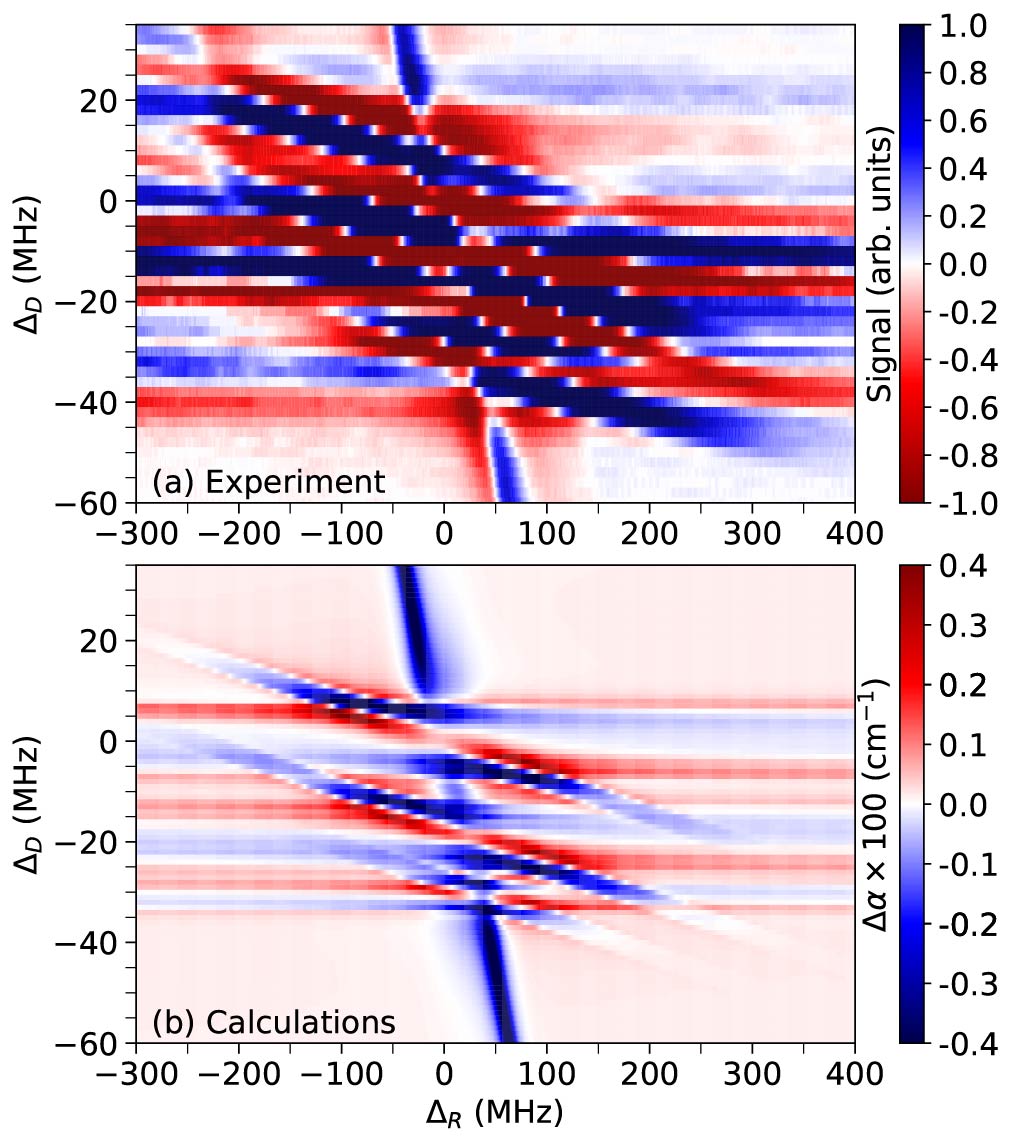}
  \caption{(a) Map of $\bar{S}(\Delta_R, \Delta_D)$ for a Rb vapor cell free of added inert gas. Coupler- and dressing-beam detunings, $\Delta_R$ and $\Delta_D$, are varied over respective ranges of 700~MHz and 95~MHz, with $\Delta_R=0, \Delta_D=0$ corresponding to the $\ket{5S_{1/2}, F = 3} \rightarrow \ket{5P_{1/2}, F' = 3}$ and $\ket{5P_{1/2}, F' = 3} \rightarrow \ket{5D_{3/2}, F'' = 4}$ double resonance. EIT corresponds with positive values (blue regions), and EIA with negative values (red regions).
  (b) Simulation results for the change of the absorption  coefficient, $\Delta \alpha$, from average. Note the inverted color bar in (b), where values $\Delta \alpha > 0$ ($\Delta \alpha < 0$) correspond to EIA (EIT). See text for detailed explanation. }
  \label{fig:usual}
\end{figure}

We first show results for a 7.5-cm-long Rb vapor cell that is free of additional inert gases and that is held at room temperature. Fig.~\ref{fig:usual}~(a) shows a map of experimental EIT spectra $\bar{S}(\Delta_R, \Delta_D)$ acquired with this cell using the described method. 
Over a range $\Delta_D \lesssim \Omega_{DP}$, the probe absorption coefficient of the atomic vapor, $\alpha$, is greatly suppressed due to two-photon EIT on the cascade $5S_{1/2}$ $\rightarrow$ $5P_{1/2}$ $\rightarrow$ $5D_{3/2}$, leading to high values of the probe transmission fraction, $T(\Delta_R, \Delta_D)$. The AC coupling of the scope in the above-described procedure removes the effect of two-photon EIT from the data, leaving only the three-photon EIT signal. The data $\bar{S}(\Delta_R, \Delta_D)$ are displayed on a linear color scale that ranges from -1 (red) for maximum three-photon EIA to 1 (blue) for maximum three-photon EIT.

The map exhibits several types of EIT and EIA features. First, there is a steep, singular EIT stripe that has a slope of -1 and that cuts across the map within the range $-30~$MHz$ \lesssim \Delta_R \lesssim 70~$MHz. In this EIT mode, referred to as EIT-mode 1, the probe laser, which has a detuning $\Delta_P \approx 0$, resonantly excites atoms that move at a velocity $v_z \approx 0$ in the probe-beam direction. The transition linewidth of 6~MHz corresponds with a velocity bandwidth $v_z \lesssim 5$~m/s of resonant atoms. The dressing and coupler lasers resonantly drive the $\ket{5P_{1/2}, F'=3} \rightarrow$ $\ket{25F_{5/2}, F'''=*}$ two-photon transition for atoms with $v_z \approx 0$, with intermediate detunings $\delta$ from the various $\ket{5D_{3/2}, F''=*}$ hyperfine states.
For $\Delta_D \gtrsim 20~$MHz or $\Delta_D \lesssim -45~$MHz in Fig.~\ref{fig:usual}~(a), the two-photon transitions $\ket{5P_{1/2}, F'=3} \rightarrow$ $\ket{25F_{5/2}, F'''=*}$ are relatively far-off-resonance from all $\ket{5D_{3/2}, F''=*}$ hyperfine levels, and the respective two-photon Rabi frequencies are $\Omega_{RP} \sim \Omega_{RD} \Omega_{DP} / (2 \delta)$, with one-photon Rabi frequencies of the Rydberg and dressing transitions, $\Omega_{RD}$ and $\Omega_{DP}$, respectively, and intermediate detunings $\delta$ depending on $F''$.

The two-photon couplings in EIT mode 1 undergo three intermediate-state resonances where $\delta \sim 0$ because there are three allowed intermediate $\ket{5D_{3/2}, F''}$ hyperfine states, $F''=4$, 3, and 2. These have respective hyperfine detunings $\delta_{HFS, F''} = 0$, $-18.4$~MHz and $-30.2$~MHz~\cite{nez1993}. In the EIT map, the three resonance points are at $\Delta_D = - \Delta_R = \delta_{HFS, F''}$. The resonance points, in combination with the strong dressing drive, result in a complex line pattern in the region $-45~$MHz$ \lesssim \Delta_D \lesssim 20$~MHz. This pattern includes EIT-mode 2, in which the probe and dressing transitions resonantly drive a
$\ket{5S_{1/2}, F=3} \rightarrow \ket{5D_{3/2}, F''}$ two-photon transition, while the coupler resonantly drives the $\ket{5D_{3/2}, F''} \rightarrow \ket{25F_{5/2}, F'''=*}$ one-photon transition. In our case it is $\Delta_P \approx 0$, and the EIT mode 2 is associated with atoms moving at a velocity of $v_z = - 18.36 \times (\Delta_D-\delta_{HFS, F''})$~(m/s)/MHz relative to the probe-beam direction (see Sec.~\ref{sec:theory}). The $\Delta_D$- and  $\Delta_R$-ranges over which EIT mode 2 is observed are limited by the width of the Maxwell velocity distribution of the Rb atoms in the vapor cell.

In Fig.~\ref{fig:usual}~(a), the EIT mode 2 gives rise to linear blue stripes with a shallow slope of $d \Delta_D/ d \Delta_R \approx -0.070$. For small $\Omega_{DP} \lesssim 5$~MHz, each $\vert 5D_{3/2}, F'' \rangle$ hyperfine state would generate a single EIT-mode 2 stripe intersecting the EIT-mode 1 stripe at $\Delta_D = - \Delta_R = \delta_{HFS, F''}$. However, in our experiment the dressing Rabi frequencies $\Omega_{DP}$ are on the order of tens of MHz.
The resultant Autler-Townes-splittings on the dressing transition tend to break the EIT mode 2 features up into two branches for each $F''$, leading to a total of up to 6 branches. As a result, near the intersection points of the steep EIT mode 1 feature (slope $d \Delta_D/ d \Delta_R \approx -1$) and the shallow EIT-2 features (slope $d \Delta_D/ d \Delta_R \approx -0.070$), in Fig.~\ref{fig:usual}~(a) we observe considerable complexity, including regions of strong EIA that break up the EIT mode 1 feature.

At certain constant values of $\Delta_D$ and large $| \Delta_R |$ Fig.~\ref{fig:usual}~(a) further exhibits horizontal bands of alternating EIT and EIA. These are likely caused by interference between the EIT modes 1 and 2, as well as their crosstalk with two-photon EIT on the three-level cascade $|5S_{1/2}$ $\rightarrow$ $|5P_{1/2} \rangle$ $\rightarrow$ $|5D_{3/2} \rangle$. 

While some features in maps such as the one displayed in Fig.~\ref{fig:usual}~(a) have qualitative explanations, detailed interpretations of these features require a simulation of our three-photon EIT system. The simulations presented next will provide further insights.

\section{Theoretical model}
\label{sec:theory}

We use a Rydberg-EIT model based on Lindblad equations, similar to that in~\cite{Thaicharoen2019}, to determine Beer's absorption coefficient, $\alpha$, for the EIT-probe beam. Since all polarizations are linear and parallel, for $^{85}$Rb the Hilbert space of coherently coupled states breaks up into sub-spaces of levels
$\{ \vert 5S_{1/2}, F=3, m_F \rangle,
\vert 5P_{1/2}, F'=3, m_F \rangle,
\vert 5D_{3/2}, F''=*, m_F \rangle,
\vert 25F_{5/2}, F'''=*, m_F \rangle \}$. The $F=2$ and $F'=2$ hyperfine states of $5S_{1/2}$ and $5P_{1/2}$, respectively, are not included because they are far-off-resonant~\cite{SteckRb85}. The quantization axis is given by the polarization direction of the laser beams. Each subspace of laser-coupled states has a fixed magnetic quantum number $m_F$ that can range from -3 to 3. The hyperfine states of $5D_{3/2}$ are normally ordered, and the splittings between the optically coupled ones are 18.42~MHz between $F''=3$ and 4, and 11.81~MHz between $F''=2$ and 3~\cite{nez1993}. The hyperfine states of the Rydberg level $25F_{5/2}$ are practically degenerate, and the optically coupled ones have hyperfine quantum numbers $F'''=1$ to 5. Therefore, the EIT simulation includes up to 10 levels, which are coupled by up to 13 allowed single-photon couplings, dependent on $m_F$. The Rabi frequencies for the couplings are determined as described in~\cite{LeKien2013}, using reduced matrix elements from~\cite{safronova2004, safronova2011}. For the radial matrix element of the Rydberg transition we use a private code. The two-photon couplings from $\vert 5P_{1/2}, F'=3, m_F \rangle$ to any $\vert 25F_{5/2}, F''', m_F \rangle \}$ are generally coherent sums over multiple intermediate states $\vert 5D_{3/2}, F''=*, m_F \rangle$.

Sub-spaces for different $m_F$ become coupled through the decay channels, leading to optical pumping. With optical pumping rates on the order of $10^7$~s$^{-1}$ and atom-field interaction times $\lesssim 1~\mu$s (probe beam size divided by thermal velocity), only on the order of 5 probe photons are scattered by each atom passing through the EIT probe region. Hence, optical pumping effects are deemed fairly weak. Therefore, in our numerical solutions of the Lindblad equation we treat the $m_F$ manifolds separately and assume $m_F$-preserving decays. The simulation results for the individual $m_F$ are averaged over the allowed $m_F$-range, $m_F = -3, ..., 3$, using equal weights for the $m_F$. The simulations only need to be conducted for $m_F =1$, 2 and 3 because the results depend only on $|m_F|$, and because atoms in $m_F=0$ do not absorb $\pi$-polarized EIT-probe light on the chosen $F=3$ to $F'=3$ EIT probe transition. Further, different atom velocities result in velocity-dependent Doppler shifts of all three laser fields, necessitating separate numerical solution of the Lindblad equation on a sufficiently fine grid of the atom velocity in the laser beam direction, $v_z$. The results are velocity-averaged with a weighting given by the one-dimensional Maxwell velocity distribution of the atoms in the cell~\cite{Thaicharoen2019}. 

The decay rates in our simulations are set as follows. The Rydberg-state decay rate is sufficiently small that it can be modeled by an effective recycling of Rydberg atoms into the $5S_{1/2}$ ground-state population. Physically, the recycling amounts to Rydberg atoms leaving and ground-state atoms entering the EIT probe region due to thermal motion in the 300-Kelvin cell. In our model, we account for the atom recycling by introducing decays from the five populated Rydberg hyperfine levels, $\vert 25F_{5/2}, F''', m_F \rangle$ with $F'''=1$ to 5, into $\vert 5S_{1/2}, F=3, m_F \rangle$ at a decay rate of $2 \pi \times 0.1$~MHz. The Rb $5D_{3/2}$ lifetime has been measured to be 246~ns~\cite{Sheng2008}, corresponding to a decay rate of about $2 \pi \times 0.7$~MHz. The $5D_{3/2}$-decay proceeds predominantly through the rapidly-decaying $5P_J$ and $6P_J$ states, with $5P_{1/2}$ being the dominant intermediate state in the decay chain. In our model the $5D_{3/2}$-decay is approximated by a single decay channel from $5D_{3/2}$ to $5P_{1/2}$ with a decay rate of $2 \pi \times 0.8$~MHz. The $5P_{1/2}$ to $5S_{1/2}$ decay rate is set to $2 \pi \times 6$~MHz~\cite{SteckRb85}.

It is noted that the absence of $m_F$-crosstalk in solving the Lindblad equation is a critical simplification because it reduces the Hilbert space dimension to only 10 states, corresponding to 100 real-valued elements of the density operator. The state space of all laser-coupled states has a dimension of 49, and the global space of laser-coupled states plus states populated via decays is even larger, corresponding to density operators with many thousands of matrix elements (which would have to be solved for on a grid of several thousand velocities in the thermal atom sample). The Lindblad equation describing Rydberg-EIT dynamics in large Hilbert spaces can be solved using a stochastic method that utilizes ensembles of wave-function ``trajectories''~\cite{Xue2019, Zhang2018}. We do not believe that this approach would reveal additional insights that would become observable at the accuracy level of our present experimental study. However, in the future, the modeling of refined measurements may warrant simulations based on a quantum trajectory approach in the global Hilbert space of the system.

Fig.~\ref{fig:usual}~(b) shows the results of a simulation performed as described above. 
The simulation is for radial Rabi frequencies of 24~MHz, 80~MHz, and 20~MHz for the probe, dressing and coupler beams, respectively. All Rabi frequencies are reduced by angular factors that are mostly $\lesssim 0.5$ and that depend on all upper- and lower-state angular quantum numbers.
In accordance with the experimental procedure, we display 
\[ \Delta \alpha (\Delta_R, \Delta_D) = \alpha (\Delta_R, \Delta_D) - \langle \alpha (\Delta_R, \Delta_D) \rangle_{\Delta_R} \quad, \]
where $\alpha (\Delta_R, \Delta_D)$  is the Beer's absorption coefficient computed on a two-dimensional grid of coupler and dressing beam detunings with a step size of 1~MHz, and the average $\langle \rangle$ is taken only over the coupler detuning, $\Delta_R$, with $\Delta_D$ held fixed.

The computation reproduces the experimental observations quite well. We observe both EIT modes 1 and 2 discussed in Sec.~\ref{sec:without}. The computations confirm that the ``fishbone" pattern of the EIT mode 2 features is due to a combination of the resolved $5D_{3/2}$ hyperfine structure and Autler-Townes couplings introduced by the dressing-beam Rabi frequencies, which are tens of MHz. Due to the small size of the probe beam relative to both the dressing and the coupler beams ($\approx 80~\mu$m versus $\gtrsim 250~\mu$m), inhomogeneous broadening due to spatial variations of the dressing- and coupler-beam Rabi frequencies across the EIT-probe beam's sampling volume is insignificant (as was seen in numerical modeling not presented). However, the dressing-beam Rabi frequencies depend strongly on $F''$ and $m_F$, leading to different Autler-Townes splittings for the $\vert 5D_{3/2}, F'', m_F \rangle$ hyperfine levels. These result in six broadened branches of EIT mode 2 that intersect with the stripe for the EIT mode 1. The EIT mode 2 branches have slopes of $\approx -0.070$ at $\Delta_R$-detunings far from the EIT mode 1 stripe.  Finally, the pattern of horizontal, alternating EIT and EIA stripes, observed in the experimental data in Fig.~\ref{fig:cases}~(a) at large $|\Delta_R|$, is also qualitatively reproduced in Fig.~\ref{fig:cases}~(b). Deviations in the details between the experiment and the simulation are attributed, in part, to environmental magnetic fields, which are not shielded, laser polarization imperfections, and optical-pumping effects~\cite{Moon:08,he2013,Zhang2018,sujap2022}, which are neglected here.

\section{Effects of an added inert gas}
\label{sec:with}

\begin{figure}[htb]
 \centering
  \includegraphics[width=0.47\textwidth]{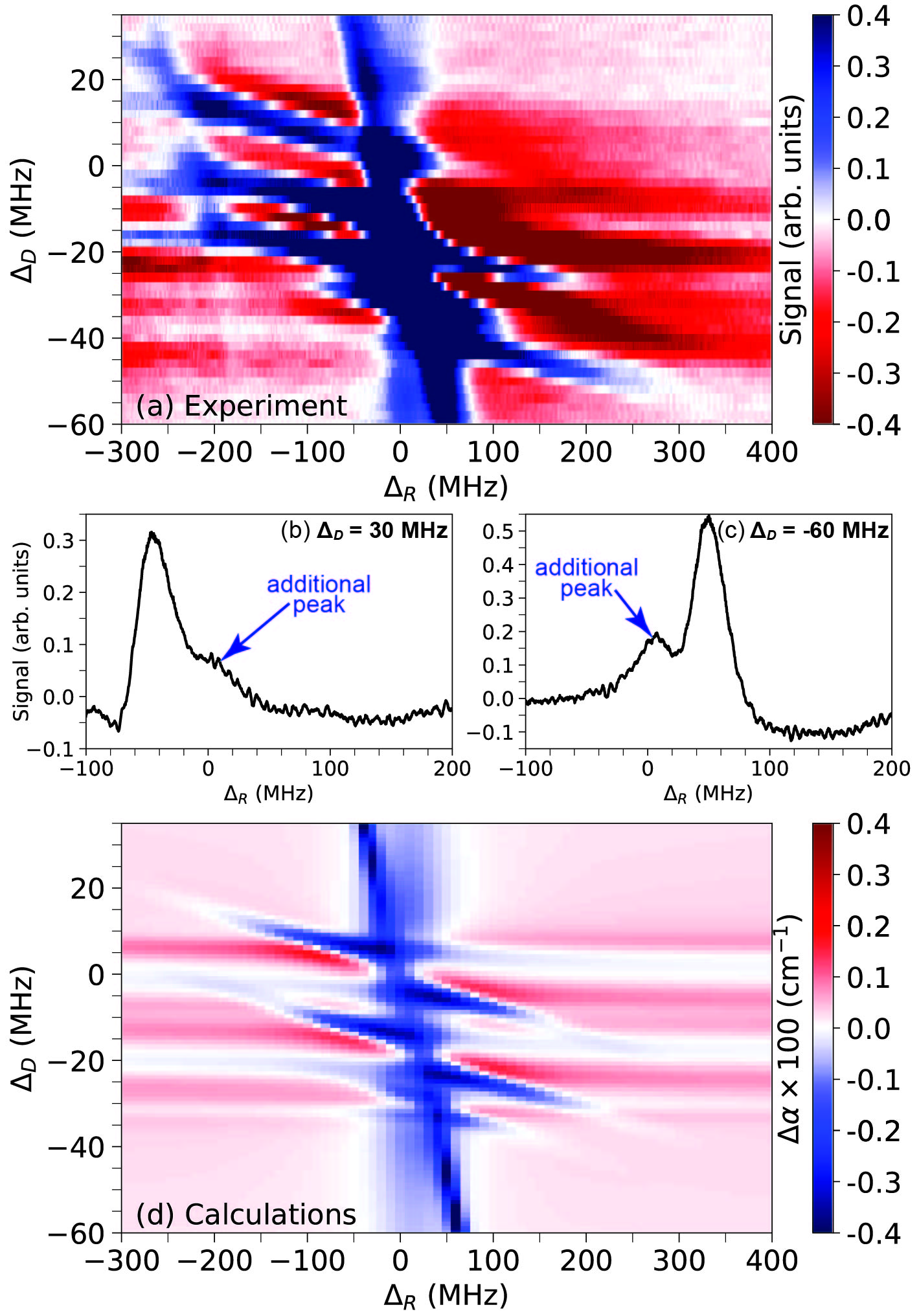}
  \caption{(a) Same as Fig.~\ref{fig:usual}~(a), but for a Rb vapor cell with an added Ar gas at 50~mTorr pressure. Two individual coupler-laser scans at $\Delta_D = $~30 and~-60~MHz are shown in (b) and (c), respectively. Panel (d) shows the result of a simulation of $\Delta \alpha$ analogous to Fig.~\ref{fig:usual}~(b), but with hyperfine mixing in $5D_{3/2}$. See text for details.} 
  \label{fig:buffer}
\end{figure}

To explore the utility of Rydberg-EIT as a diagnostic tool to measure plasma electric fields, an inert gas that is sufficiently dense to support a low-pressure plasma must be introduced. In previous work on two-photon Rydberg EIT, it was found that inert-gas collisions can reduce the visibility of EIT lines and add additional features~\cite{Thaicharoen2024}. 
In the following, we investigate the effects of low-pressure argon (Ar) gas in considerably richer three-photon EIT spectra. In the experiment, we replace the 7.5~cm long Ar-free vapor cell with a 4~cm long cell containing a mixture of Rb vapor and 50~mTorr of Ar.

Experimental three-photon EIT spectra taken in the presence of 50~mTorr of Ar are shown in Fig.~\ref{fig:buffer}~(a). While spectra without [Fig.~\ref{fig:usual}~(a)] and with 50~mTorr of Ar [Fig.~\ref{fig:buffer}~(a)] appear qualitatively similar, there are several important differences.
First, the Ar gas reduces the overall signal strength of the EIT 1 and EIT 2 modes (note the color scales are different in the figures). The EIT mode 2 features remaining in the presence of the Ar gas are most prominent in the upper-left and lower-right segments of the plot.
In the intersection region between the EIT mode 1 and EIT mode 2 features, the EIT mode 1 line becomes contiguous and is not broken up by regions of strong EIA, as is the case in Fig.~\ref{fig:usual}~(a). 
The visibility of the pattern of alternating horizontal EIT and EIA stripes at large $\vert \Delta_R \vert$ and $\Delta_D \sim 0$ is reduced in Fig.~\ref{fig:buffer} relative to Fig.~\ref{fig:usual}. Notably, the Ar gas introduces an additional EIT feature at near-zero coupler detuning, $\Delta_R \approx 0$. This feature, which we refer to as EIT mode 3, becomes visible for $\Delta_D \gtrsim 10$~MHz and $\Delta_D \lesssim -40$~MHz, and its position on the $\Delta_R$-axis is independent of $\Delta_D$. Two instances of the EIT mode 3 in coupler-laser scans at selected values of $\Delta_D$ are shown in Figs.~\ref{fig:buffer}~(b) and~(c), where they are labeled as ``additional peak''.

It is noteworthy that the Ar gas red-shifts the Rydberg lines by $\approx$~-20~MHz~\cite{dash2024, lei2024}. Further, results qualitatively similar to those in the present work were obtained with 500~mTorr of Ar (not shown). The overall signal diminishes in strength and broadens with increasing pressure, preventing a meaningful analysis at 500~mTorr.

\section{Modeling of inert-gas effects}
\label{sec:modeling}

In the following, we introduce hyperfine collisions of the Rb states in our model to obtain insight into which interactions may explain the changes between Fig.~\ref{fig:buffer}~(a) and Fig.~\ref{fig:usual}~(a).
For simplicity, we first use a 4-level model with a single $5D_{3/2}$ and $25F_{5/2}$ state. Hyperfine-changing collisions and depolarization are modeled via a dephasing term that selectively acts upon one of the four involved states. Comparing the resultant simulated EIT maps with 
Fig.~\ref{fig:buffer}~(a) then allows us to assess which type of interaction is the most consistent with the experimental data. To clarify our notation for decay rates, assume that a level $\vert n \rangle$ decays to level $\vert m \rangle$ with a population decay rate $\Gamma_{mn}$, and that level $\vert n \rangle$  exhibits additional dephasing ({\sl{i.e.}} coherence decay without population decay) according to a rate $\gamma_n$. In our model, these processes
would be described by damping of the form
\[ \dot{\rho}_{mn} = - \frac{(\Gamma_{mn} + \gamma_{n})}{2} \rho_{mn}.\]

\begin{figure*}[htb]
 \centering
  \includegraphics[width=\textwidth]{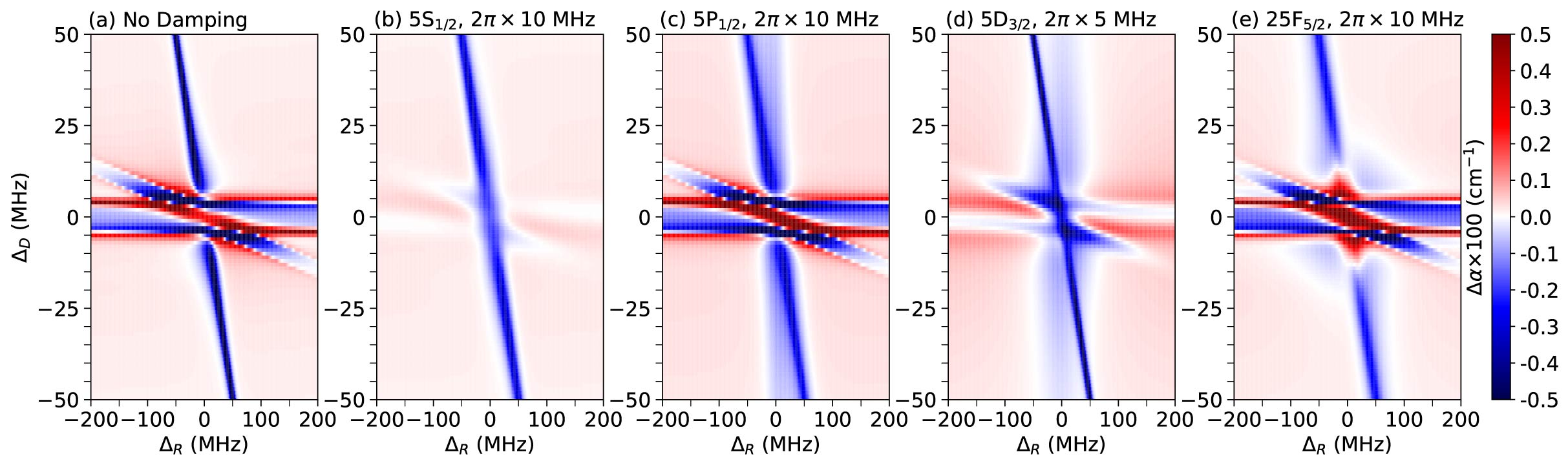}
  \caption{Effects of collisional level dephasing of different states involved in the process of three-photon Rydberg-EIT using the cascade the $5S_{1/2} \rightarrow$ $5P_{1/2} \rightarrow$ $ 5D_{3/2} \rightarrow $ $ 25F_{5/2}$. In panel (a) there is no collisional level dephasing. In panels (b) - (e), collisional level dephasing is applied to the indicated states at the rates shown.}
  \label{fig:cases}
\end{figure*}

In Figs.~\ref{fig:cases}~(a)-(e) we use probe, dressing and coupler Rabi frequencies of $2 \pi \times$5~MHz, $2 \pi \times$20~MHz and $2 \pi \times$10~MHz, respectively, which are on the order of the magnitudes used in the experiment. We also use the above population decay rates of $2 \pi \times$6~MHz from $5P_{1/2}$ to $5S_{1/2}$,  $2 \pi \times$0.8~MHz from $5D_{3/2}$ to $5P_{1/2}$, and a Rydberg-atom recycling rate of $2 \pi \times$0.1~MHz from $25F_{5/2}$ to $5S_{1/2}$. In Fig.~\ref{fig:cases}~(a) 
there is no collisional level dephasing, while in Figs.~\ref{fig:cases}~(b)-(e)  
we add collisional dephasing at rates of $2\pi \times$10~MHz on $5S_{1/2}$, $2\pi \times$10~MHz on $5P_{1/2}$, $2\pi \times$5~MHz on  $5D_{3/2}$, and $2\pi \times$10~MHz on  $25F_{5/2}$, respectively. Only one of the four levels at a time is subjected to collisional dephasing. It is seen in Figs.~\ref{fig:cases} that the four types of dephasing reduce the signal strength of the remaining EIT/EIA features by quite different degrees, allowing us to isolate potential causes for our experimental observation. 

The simulation result in Fig.~\ref{fig:cases}~(b) eliminates $5S_{1/2}$ level dephasing as a significant cause of our experimental findings because it fails to produce the EIT mode 3, and because it almost entirely eliminates the EIT mode 2. 

Next, from Fig.~\ref{fig:cases}~(c)
we conclude that $5P_{1/2}$ level dephasing may be present at some level, because it raises the EIT  mode 3 feature. However, $5P_{1/2}$ dephasing leaves the EIT mode 2 feature too strong relative to the EIT mode 1 feature. Further, the EIT mode 1 feature still remains broken up by a region of strong EIA near $\Delta_D = 0$. Moreover, $5P_{1/2}$ dephasing does not sufficiently reduce the visibility of the pattern of alternating horizontal EIT and EIA stripes at large $\vert \Delta_R \vert$. 

In Fig.~\ref{fig:cases}~(d) we see that the $5D_{3/2}$ collisional dephasing raises the EIT mode 3 feature and attenuates both EIT modes 1 and 2 by similar amounts, while preferentially preserving the upper-left and lower-right branches of the EIT mode 2 features. Moreover, the $5D_{3/2}$ dephasing changes the EIT mode 1 into a contiguous feature that is not broken up by an EIA region near $\Delta_D = 0$. Further, the visibility of the pattern of alternating horizontal EIT and EIA stripes at large $\vert \Delta_R \vert$ is reduced but not entirely washed out. Since the simulated effects of the $5D_{3/2}$ level damping largely conform with our experimental observations, we believe that processes consistent with the $5D_{3/2}$ collisional dephasing, such as $5D_{3/2}$ hyperfine-changing collisions and depolarization, are the most prevalent in our experiment. 

Finally, based on Fig.~\ref{fig:cases}~(e), Rydberg-state dephasing is deemed unimportant because it does not raise the EIT mode 3, it does not significantly attenuate the EIT mode 2 and the pattern of alternating horizontal EIT and EIA stripes at large $\vert \Delta_R \vert$, and it enhances the regions of EIA that break up the EIT mode 1 feature instead of eliminating them.  

From Fig.~\ref{fig:cases} we conclude that inert-gas-induced $5D_{3/2}$ hyperfine dephasing should explain most of the differences between the measurements in Figs.~\ref{fig:buffer} and~\ref{fig:usual}. As to EIT mode 3, we argue that the level broadening of $5D_{3/2}$ caused by the $5D_{3/2}$ hyperfine dephasing leads to a population of near-zero-velocity atoms in $5D_{3/2}$ even at large detunings $\Delta_D$, generated via resonant excitation by the weak EIT probe beam and off-resonant excitation by the more intense EIT dressing beam. Due to their near-zero velocity, these atoms become resonantly excited into the Rydberg state at $\Delta_R \approx 0$, leading to a reduced population in $5P_{1/2}$. The latter, in turn, manifests as the EIT mode 3 band, which occurs at $\Delta_R \approx 0$ and stretches over a wide range of $\Delta_D$. 

The above conclusions are reinforced by a simulation result in Fig.~\ref{fig:buffer}~(d), where we have used our 10-level model and added collisional hyperfine-changing transitions between all $\vert 5D_{3/2}, F''=*\rangle$ hyperfine states at transition rates of $\Gamma_{mn} = 2 \pi \times 1.5$~MHz, with both $m$ and $n$ counting from $F''=2$ to $F''=4$ and $m \ne n$. We observe good qualitative agreement with the experimental result in Fig.~\ref{fig:buffer}~(a). The full 10-level simulations also confirm that the effects of hyperfine level dephasing and hyperfine-changing transitions in the EIT spectrum are largely the same.

\section{Discussion and Conclusion}
\label{sec:disc}

We have investigated hyperfine effects of the $5D_{3/2}$ level in Rb 3-photon EIT 
on the cascade $5S_{1/2} \rightarrow5P_{1/2} \rightarrow 5D_{3/2} \rightarrow 25F_{5/2}$ in a Rb vapor cell at 300~K. We found two modes of EIT as well as a pattern of EIT and EIA features that are related to the $5D_{3/2}$ hyperfine structure. Adding a low-pressure Ar gas at 50~mTorr caused significant changes, including the third EIT mode, which we mostly attribute to hyperfine-changing collisions and depolarization of the $5D_{3/2}$ level, the highest excited state of the cascade system that is not a Rydberg state. The experimental observations were reproduced at a good qualitative level by simulations, which were also essential in isolating the causes for the observed inert-gas-induced changes.      

The measured and simulated results presented in Figs.~\ref{fig:usual} and~\ref{fig:buffer} differ in some details. The differences may, in part, be due to environmental magnetic fields that were not shielded in the experiment. Also, the three laser beams may have had polarization errors of up to $\sim 10\%$ due to the use of polarization-maintaining fibers. More fundamentally, in our model we have neglected optical pumping, which could be addressed using a stochastic method based on ensembles of wave-function ``trajectories''~\cite{Xue2019, Zhang2018}. Further, in our present work we have computed steady-state solutions of the density operator. In the case of an added inert gas, the atoms are subject to spatial diffusion dynamics and velocity-changing collisions, which may prolong the EIT/EIA time constants of the vapor during the laser scans~\cite{dash2024, lei2024}. In future research, these topics could be explored. 

Finally, we anticipate that our observations may lead to renewed studies on interactions of alkali Rydberg atoms with inert gases, especially studies involving excited electronic states~\cite{Dubourg_1986}. Previously, line broadening and line shifts in similar systems have attracted considerable attention~\cite{brillet1980,thompson1987, Megan_2002, sargsyan2010, Zameroski_2014}. Our results point to state-dependent hyperfine collisions, revealed by additional lines in multi-photon Rydberg-EIT spectra. Future probes of collisions in Rydberg-EIT would be of interest in the context of recent advanced theoretical studies on alkali-noble-gas collisions~\cite{medvedev2018} as well as developments in modern vapor cell technologies~\cite{kitchingreview2011}. Further, many-body spin interactions in such systems are of interest in room-temperature quantum information methods~\cite{katz2022}.

\section*{Acknowledgments}
\label{sec:acknowledgments}
We would like to thank Dr. Ryan Cardman for initial contributions to the construction of the experimental setup and Bineet Dash, Dr. Nithiwadee Thaicharoen, Dr. Eric Paradis, and Dr. David A. Anderson for useful discussions. This work was supported by the U.S. Department of Energy, Office of Science, Office of Fusion Energy Sciences under award number DE-SC0023090, and NSF Grant No. PHY-2110049. A.D. acknowledges support from a Rackham Predoctoral Fellowship at the University of Michigan.

\bibliography{bibliography.bib}

\end{document}